\pdfoutput=1 % only if pdf/png/jpg images are used
\documentclass{JINST}
\usepackage{textcomp}
\usepackage{graphicx}
\usepackage{siunitx}
\usepackage{xcolor}
\let\normalcolor\relax
\title{The Pixel-TPC: first results from an 8-InGrid module}

\author{M. Lupberger$^a$\thanks{Corresponding author.}~
for the LCTPC collaboration\\
\llap{$^a$}University of Bonn,\\
  Bonn, Germany\\

E-mail: \email{lupberger@physik.uni-bonn.de}}

\abstract{An eight chip InGrid module has been tested as readout of a prototype time projection chamber. The construction of the modules as well as first preliminary results are presented. The InGrids, a Micromegas like amplification structure with a pixel ASIC, have been produced on wafer-level in a photolithographic process. For the measurements a newly developed readout system based on the Scalable Readout System was used.}


\begin{document}

\section{Introduction}\label{sec:intro}
The time projection chamber (TPC) was invented in 1976~\cite{bib0} and is still used in today's particle physics experiments. Also for the International Large Detector (ILD)~\cite{ILD}, one of the two detectors foreseen for the future International Linear Collider (ILC), a TPC is planed as the major tracking detector. The most recent technology to equip the TPC-endplate are micropattern gaseous detectors (MPGDs).
One of the concepts are Micromegas~\cite{bib1} which intrinsically come with a high granularity that is given by the distance between the holes in the grid. To reflect this from the readout side an ASIC, the Timepix chip~\cite{bib2}, with a pixel pitch of \SI{55}{\micro\meter} is used in our experiments. Such an ASIC with a post processed grid on top is called InGrid~\cite{chefdeville}. It is our goal to demonstrate that such a type of detector can be used to read out the ILD TPC.
\subsection{Motivation}\label{sec:motiv}
Micro-structuring of semiconductor devices was a breakthrough regarding the tracking capabilities of silicon detectors. In most of the particle physics experiments, strip and pixel detectors are used as vertex and tracking detectors in the central region. For gaseous detectors, micro-structuring of the readout anode led to a similar revolution with MPGDs replacing wires. The dimensions of the amplification structure could be reduced from the millimetre scale down to some \SI{10}{\micro\meter}. However, the dimension of a readout pad, that is used at many MPGDs, still does not match the high granularity. The amount of electronics involved in processing pad signals prevents to move to smaller pad sizes. Using ASICs with integrated electronics and about \SI{10}{\micro\meter} pixel size could be a solution to overcome this limitations.\\
In the case of thin planar drift detectors aiming for a high transverse spatial resolution $\mathrm{\sigma_{xy}} $, the limiting factor is the transverse diffusion constant $\mathrm{D_T}$ of primary electrons in the gas volume. Assuming an electron cloud from a primary ionisation with $O(10)$ electrons, the cloud will have a size $O(30$)\,\SI{}{\micro\meter} after one millimeter of drift in a gas with $\mathrm{D_T} = O(100)$\,\SI[per-mode=symbol]{}{\micro\meter\per\sqrt{\centi\meter}}. Hence, the best possible single point resolution is $O(10)$\,\SI{}{\micro\meter}, which is comparable to a silicon detector.\\
For long drift distances as in case of a TPC, the benefits of a fine-grained readout plane are the detailed visualisation of $\mathrm{\delta}$ electrons, the double-track resolution, the direct $\mathrm{dE/dx}$ measurement by cluster counting (see section~\ref{sec:InGrid}) and almost no track angle effects, as the pixels are quadratic and not rectangular. The data taken in the test beam campaign discussed later will be used to demonstrate these advantages.\\
As far as X-ray detection is essential, a pixel MPGD is capable to distinguish between photons, minimum ionising particles and alpha particles by means of pattern recognition, ionisation density and energy deposition. Depending on the gas, even photoelectron tracks can be recognised.\\
Because of these features,  the InGrid is a good candidate to be used in rare event searches as CAST~\cite{cast} and DARWIN~\cite{darwin}, vertex detectors~\cite{gossip} or the ILD TPC.
\subsection{History}\label{sec:hist}
The concept of combining a pixel ASIC with a MPGD was first discussed by Bellazzini and Spandre in \cite{bellazzini} and Colas et al. in \cite{ColasNikhef}. The former used a Gas Elektron Multiplier (GEM) with \SI{60}{\micro\meter} pitch in combination with a 512~pixel readout with \SI{200}{\micro\meter} pitch. Their detector was built to measure the tracks of photo electrons for X-ray polarimetry. They noted that \textit{"the real challenge with this class of detectors is the design of the read-out system which should not spoil the intrinsic performance of the device"}.\\
Colas and his colleagues used the Medipix2 chip~\cite{medipix2} in combination with a triple GEM stack or a Micromegas. An iron source and cosmic rays were used to produce primary electrons. In this publication it was already mentioned that the goal of this effort is the development of a monolithic integrated Micromegas that was called TimePixGrid at that time.\\
In a later publication~\cite{singlee} the same group reported about the possibility to detect single electrons with 90~\% efficiency with this detector. For their measurements they used tracks of cosmic particles and also recorded $\mathrm{\delta}$ electrons.\\
The next step was to investigate a technique to align the holes in the grid with the pixels of the chip and control parameters like the hole size, gap height and grid thickness. The first approach was to develop a technology to build an aluminium grid on top of a bare silicon wafer, which is reported in \cite{postprocess}.
This was followed by a detailed study~\cite{chefdeville} on the fabrication and testing of an integrated grid on a CMOS pixel chip, the Timepix chip.\\
Another important step was the first test beam campaign with combined MPGD and pixel readout which was carried out at the \SI{5}{\giga\electronvolt} electron beam of the DESY II synchrotron~\cite{bamberger}. With a setup of a triple-GEM stack and a Medipix and Timepix chip the resolution of such a detector was studied. For short drift distances, a single point resolution down to approximately \SI{25}{\micro\meter} could be achieved.\\
Up to now, MPGDs in combination with pixel readout are used only in small experiments. But a  R{\&}D proposal~\cite{AtlasNoteInGrid} was approved by the Atlas Upgrade Steering Group as meaningful R{\&}D activity.

\subsection{The Timepix chip}\label{sec:Timepix}
The ASIC used in our experiments is the Timepix chip~\cite{bib2}. It was developed in 2007 by the Medipix2 Collaboration and has a matrix of 256$\times$256 pixels, each with a size of \SI{55}{\micro\meter}$\times$\SI{55}{\micro\meter}. The sensitive area is \SI{1.4}{\centi\meter}$\times$\SI{1.4}{\centi\meter}. The input pad of each pixel is connected to a charge sensitive preamplifier and a single threshold discriminator. This analog part is then connected to the digital part of the pixel, which is driven by an external clock. Each pixel contains a 14 bit counter. The logic for this counter can be set in one of the two main modes: It can either measure charge by counting the number of clock cycles the discriminator signal is over the threshold (time over threshold, TOT) or measure the arrival time with respect to a trigger signal. This can be achieved by counting the number of clock cycles from the moment when the signal exceeds the threshold until the end of a shutter window opened by the trigger. By knowing the length of the shutter window and the trigger delay, one can calculate the arrival time. The TOT mode needs a calibration to transform the measured clock cycles to the charge that generated the signal in the preamplifier. An injection capacitor in every pixel can be used for a calibration with well defined input charge by test pulses.\\
After each shutter window the complete pixel matrix needs to be read out, that is 917504 bit. The Timepix chip is designed to be operated with a maximum external clock frequency of \SI{200}{\mega\hertz}. This results in a readout rate of at most \SI{218}{\hertz}, if the shutter length is short compared to the readout time. For a typical setup, the external clock frequency is about \SI{50}{\mega\hertz} and the shutter length is in the order of \SI{1}{\milli\second}, resulting in a readout rate of \SI{50}{\hertz}. Data processing and transmission has not been taken into account in this calculations. Chips can be connected in a daisy chain, where the data is forwarded from one chip to the next. The maximum readout rate will in this case be divided by the number chips in the chain.
\subsection{The InGrid}\label{sec:InGrid}
\begin{figure}[tbp] % figures (and tables) should go top or bottom of
                    % the page where they are first cited or in
                    % subsequent pages
\centering
\includegraphics[width=.8\textwidth]{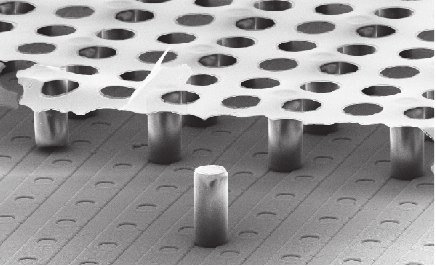}
\caption{SEM image of an InGrid with partly removed Grid made by the IZM Berlin. The height of a pillar is \SI{50}{\micro\meter}.} 
\label{fig:InGrid2}
\end{figure}

An InGrid is a special type of Micromegas. In a photolithographic process, as described in \cite{postprocess} and \cite{chefdeville}, the grid is produced on the Timepix chip and the holes of the grid are aligned to the pixels. The grid is set on a positive potential such that primary electrons, originating from ionisations above the grid, are accelerated when they enter a hole. If the electric field is large enough an avalanche of secondary electrons is created and a signal can be registered in the pixel underneath the hole. The chip is protected against sparks by a high resistive layer of silicon nitride. A SEM image of an InGrid can be seen figure~\ref{fig:InGrid2}.\\
As the distance between grid and chip is only \SI{50}{\micro\meter}, the charge will not spread to an neighbouring pixel by diffusion. However, for large gas amplifications, the protection layer could spread the charge. As long as no more than one primary electron enters a hole and the gas amplification is high enough, each primary electron will activate one pixel. It is hence possible with this type of detector to detect primary electrons with a very high efficiency.\\
An example for a measurement, where this property is used are ionisation spectra of X-ray photons in a gas. An incoming photon ionises gas atoms and releases all its energy. The electrons ionised in this process are drifted towards the grid in a low electric field. By diffusion this charge cloud is spread, such that the probability that two or more primary electrons enter one hole of the grid decreases. By counting the number of activated pixels in one recorded frame an ionisation spectrum can be generated. The high single electron detection efficiency leads to a high energy resolution. For a given gas, the number of activated pixels is proportional to the X-ray energy.
Using the same method on the ionisation density of a charged particle traversing a gas, the energy loss per track length ($\mathrm{dE/dx}$) can be directly measured by counting the number of activated pixels per track length. This can only be achieved in combination with the high spatial resolution of the Timepix chip.

\section{Experimental setup}\label{sec:setup}
In the test beam campaign described here, we operated eight InGrid chips arranged in a block of 2$\times$4, a so called octoboard. The chips are glued on a carrier board that is placed inside a module made of aluminium and connected to an intermediate board. A similar module, the Octopuce~\cite{lupberger} has already been constructed as a demonstrator and was tested in a shorter campaign with moderate gas gain.
\subsection{InGrid production}
The InGrid chips used in this campaign are from the fourth wafer scale production process performed in collaboration with the University of Twente and Fraunhofer IZM Berlin \cite{thorsten}. At the very beginning, when this technology was pioneered and optimised by NIKHEF and the University of Twente only individual chips could be processed at a time. Because of the high demands by the community due to the growing surface of detectors, this production technique has been transferred to the wafer scale. Now about 100 chips can be produced in one run. The quality and performance of these chips is similar to those produced individually. In an Argon/Isobutane 95/5 gas mixture, energy resolutions of 5~\% for and $\mathrm{Fe^{55}}$ escape peak and gas gains of 10000 can be reached.
\subsection{Readout system}
\begin{figure}[tbp] % figures (and tables) should go top or bottom of
                    % the page where they are first cited or in
                    % subsequent pages
\centering
\includegraphics[width=.6\textwidth]{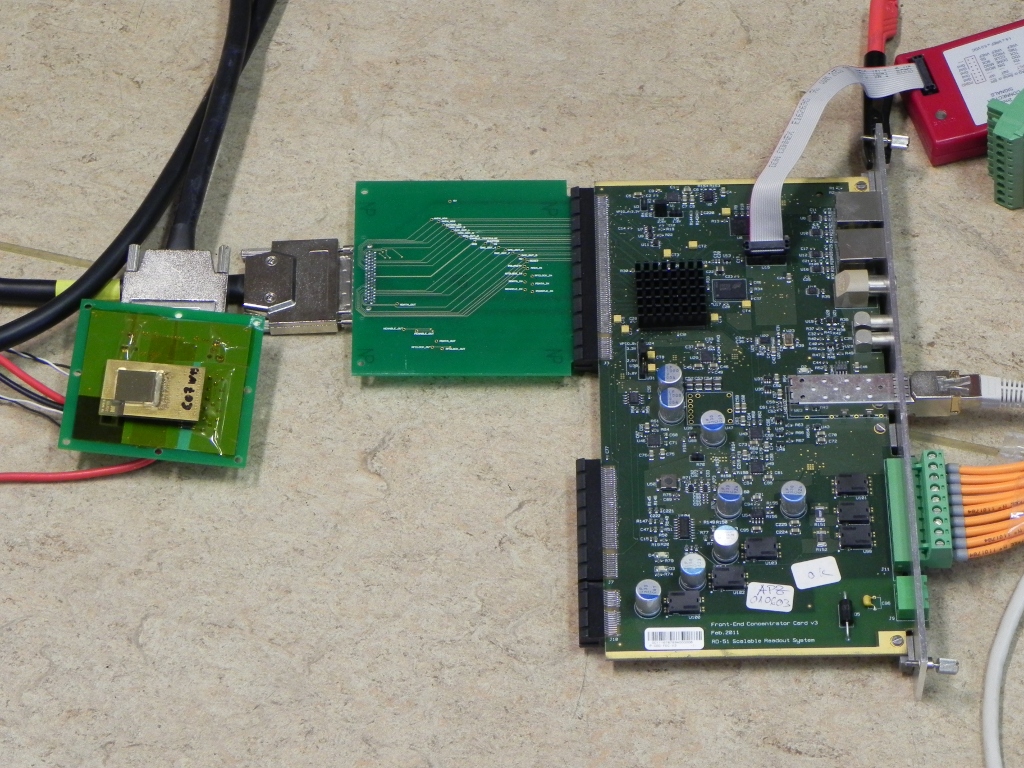}
\caption{Scalable Readout System with a single Timepix chip at the front end. From left to right: Single timepix chip on golden carrier at intermediate board connected to an adapter card (light green) at the SRS FEC.}
\label{fig:SRSFEC}
\end{figure}
Based on the Scalable Readout System (SRS)~\cite{SRS}, a new readout system for the Timepix chip has been developed. In our setup, up to eight daisy chained Timepix chips on a carrier board are plugged onto an intermediate board. This board is directly connected to a power supply for the chips and to the SRS with a VHDCI cable. An adapter card with a connector for this cable was designed to be plugged to the front end card (FEC), see figure~\ref{fig:SRSFEC}. For the FPGA on the FEC, a dedicated firmware has been developed following a first approach described in \cite{zamrowski}. The data from the chips are processed in the FPGA and sent via Gbit ethernet to a PC. For the computer, a C++ based data acquisition software was written.
\paragraph{Current status} All the functionality for data taking is implemented in software and firmware. The basic commands like reset, setting the pixel matrix, reading out the data, setting the DACs, opening or closing the shutter are performed by the FPGA. By executing combinations of these commands from the software threshold equalisations, calibrations, data taking, etc. can be done. For calibration, external test pulses are used. In the future, a multiplexer already present at the intermediate board will be used for an automatic calibration. If only one or a few chips have to be read out, a smaller system based on a Xilinx ML605 Evaluation board is also available and will be used in the CAST experiment.
\subsubsection{Requirements and solutions}
For large area detectors a modular readout system is necessary. The decision was made to use the SRS from RD51 at CERN, as with this system several FECs can be combined by an ethernet switch. Dedicated hardware was designed to support the modularity. The FPGA can be programmed by the user according to his needs. For example, zero suppression and parallelised data management have been implemented to reach the maximum theoretical readout speed of up to \SI{100}{\hertz} for a single chip.
\subsubsection{FPGA Firmware}
\begin{figure}[tbp] % figures (and tables) should go top or bottom of
                    % the page where they are first cited or in
                    % subsequent pages
\centering
\includegraphics[width=.8\textwidth]{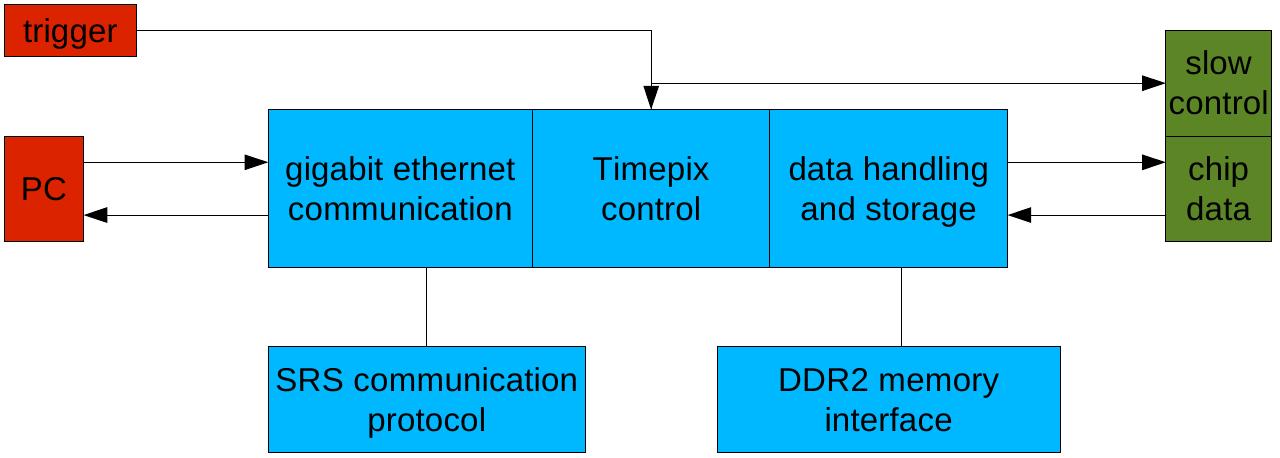}
\caption{Simplified schematic of the FPGA firmware used for the SRS Timepix readout.}
\label{fig:Firmware}
\end{figure}

The FPGA on the current SRS~FEC is a Xilinx~Virtex~5~vlx50t and will be updated to a Virtex~6 in the next version. The task of the firmware operating on this FPGA is on the one hand to control the chips and on the other hand to read out the data and transmit it to the PC as fast as possible. The code consists of several modules and is mainly written in VHDL. The ethernet communication is provided by the Xilinx Ethernet Media Access Control~(EMAC) in combination with some SRS common code. These modules communicate with the Timepix control module that transforms the ethernet commands to the command signals the chips need and vice versa. Another module is responsible for data handling and temporary data storage  during zero suppression. To extend these features, the DDR2 memory of the SRS is implemented using the Memory Interface Generator~(MIG) together with a memory control module. A simplified schematic view of the FPGA firmware can be seen in figure~\ref{fig:Firmware}.

\subsection{Test beam setup}
\begin{figure}[tbp] % figures (and tables) should go top or bottom of
                    % the page where they are first cited or in
                    % subsequent pages
\centering
\includegraphics[width=.6\textwidth]{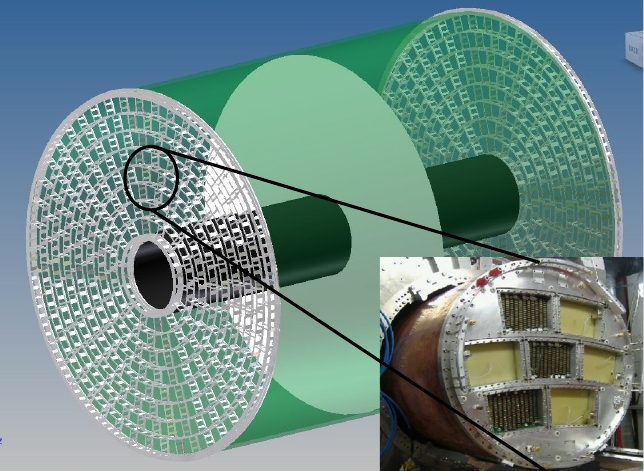}
\caption{Illustration of the ILD TPC and image of the LP endplate.}
\label{fig:LP}
\end{figure}

\begin{figure}[tbp] % figures (and tables) should go top or bottom of
                    % the page where they are first cited or in
                    % subsequent pages
\centering
\includegraphics[width=.6\textwidth]{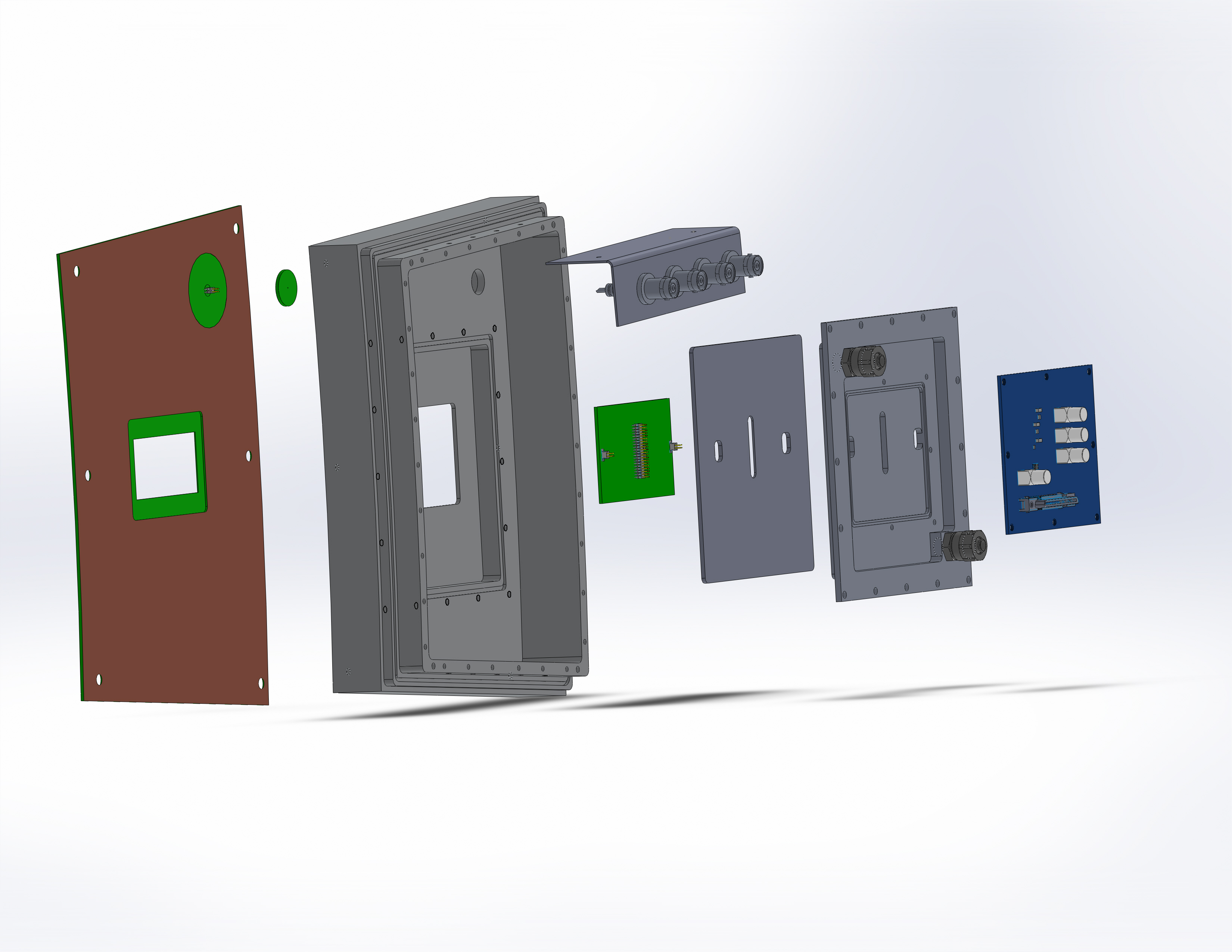}
\caption{Exploded view of the InGrid module.}
\label{fig:InGridMod}
\end{figure}

The goal of this test beam was to record tracks of ionising particles with two modules for a complete testbeam campaign and to demonstrate the functionality of the SRS system. The detectors had to show a stable behaviour at a gain with high single electron detection efficiency. Data to study the properties of pixelated readout had to be taken, as for example different track angles, momenta and z-positions. Within the data analysis this data can be used to investigate the track angular effect, $\mathrm{dE/dx}$ resolution and transverse spatial resolution.\\
For data taking, the large prototype (LP)~\cite{LP} of the linear collider TPC collaboration (LCTPC) was used. It is located in test beam area T24/1 at the DESY~II synchroton. During the 16-day campaign in spring~2013 also data with magnetic field was recorded, as the LP is inserted in a \SI{1}{\tesla} superconducting magnet called PCMAG. The TPC fieldcage has an inner diameter of \SI{72}{\centi\meter} and a length of \SI{56,76}{\centi\meter}. The anode endplate can host up to seven modules and resembles a segment of the ILD TPC endplate as shown in figure~\ref{fig:LP}. The SRS operated the Timepix chip with a clock frequency of \SI{40}{\mega\hertz} and a readout rate of \SI{2,5}{\hertz}, which is not the maximum speed for a chain of eight chips. The chips was set in the mode to measure the arrival time of the charge. With this information and the drift velocity inside the gas volume the z-position of the primary ionisation can be reconstructed. The z-axis is the axis along the TPC volume, while the x-y-plane is equivalent to the readout plane. For gas amplification in the InGrid the grid voltage was set to \SI{350}{\volt} which resembles to a gas gain of about 6000.
The trigger for the system came from the coincidence signal of two scintillators in front of the TPC. The logic for recording a frame of data was implemented in the FPGA firmware and defines a shutter window within which the chip is sensitive. The shutter window was set such that the whole TPC volume is read out. More than 2~million tracks from electrons with up to \SI{6}{\giga\electronvolt} in T2K gas (95~\% Ar, 3~\% CF${_4}$, 2\% iC$_4$H$_{10}$) have been recorded. This gas has a drift velocity of \SI[per-mode=symbol]{74}{\milli\meter\per\micro\second}. As the \SI{40}{\mega\hertz} are also the sampling frequency for the charge arrival time measurement, the resolution in z-direction can not be better than $\mathrm{\sigma_{z,min} = \frac{\SI[per-mode=symbol]{74}{\milli\meter\per\micro\second}}{\sqrt{12} \cdot 40 MHz} =} \SI{534}{\micro\meter}$.\\
The campaign included measurements at $\mathrm{B} = \SI{0}{\tesla}$ and $\mathrm{B} = \SI{1}{\tesla}$ with two different drift fields: \SI[per-mode=symbol]{230}{\volt\per\centi\meter} and \SI[per-mode=symbol]{130}{\volt\per\centi\meter} respectively. The z-position and the angle of the beam with respect to the endplate was varied as well as the electron momentum. Two different readout modules were used, one with a triple GEM amplification structure in combination with an octoboard of unprocessed Timepix chips and another one with InGrids. Analysis of GEM data has not started yet, hence, we will focus on the InGrid module.\\
In figure~\ref{fig:InGridMod} an exploded view of the module can be seen. It is made of several aluminium parts and PCBs. The outermost part on the right hand side is the intermediate board (blue). It is connected to the chip carrier board (green) by a 40 pin connector for power, control and data signals. Additionally, there are two 4-pin connectors for the grid high voltage. In between the two PCBs there is a cooling structure made of aluminium \cite{menzen}. These four parts can be plugged from the back side into the aluminium frame. The innermost part facing the gas volume of the TPC is the anode plate that also overlaps the InGrid edges in order to minimise field distortions. As the InGrid is geometrically on a different z-position than the anode plate, the electrical potential needs to be adjusted. The setting of the correct potential has been studied first in the test beam. Field distortions at the edges of the board cause primary electrons to be focussed to the anode plate or the center of the chip. If the complete surface of the chip is illuminated in an integrated image of many frames (occupancy), then the field distortions are minimised. Hence, the voltage difference between grid and anode was varied until the occupancy of the octoboards has been optimised.\\

\subsection{Preliminary analysis}
\begin{figure}[tbp] % figures (and tables) should go top or bottom of
                    % the page where they are first cited or in
                    % subsequent pages
\centering
\includegraphics[width=.8\textwidth]{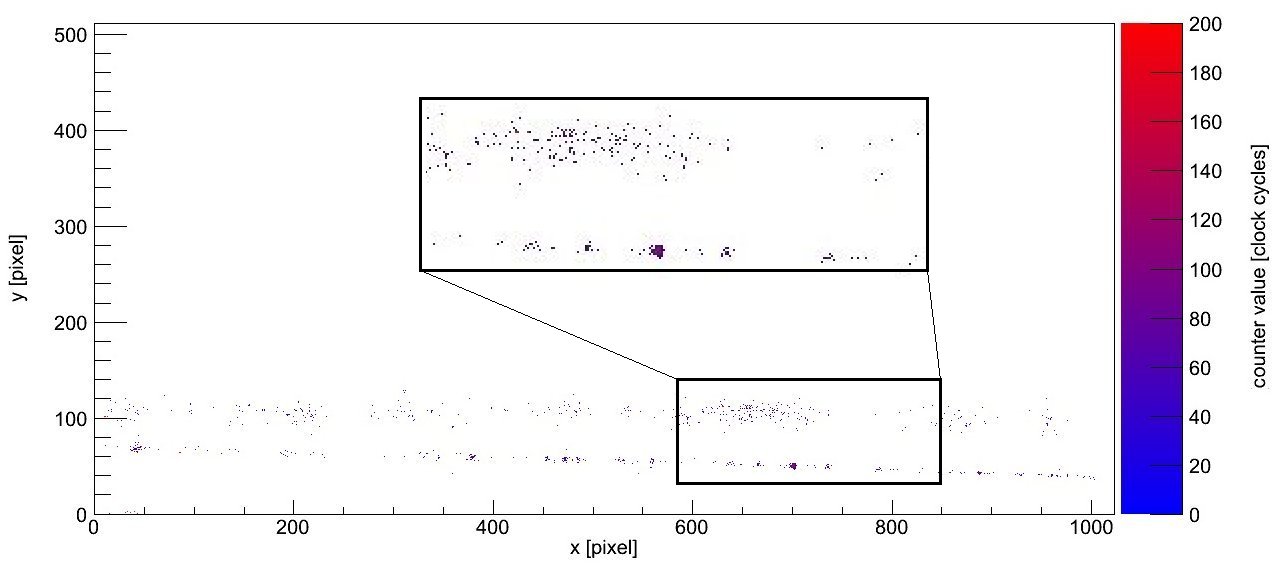}
\caption{Online event display of double track event from the InGrid module. The two tracks stem from different z-positions as they have a different transverse diffusion width.}
\label{fig:eventdisplay1}
\end{figure}

\begin{figure}[tbp] % figures (and tables) should go top or bottom of
                    % the page where they are first cited or in
                    % subsequent pages
\centering
\includegraphics[width=.8\textwidth]{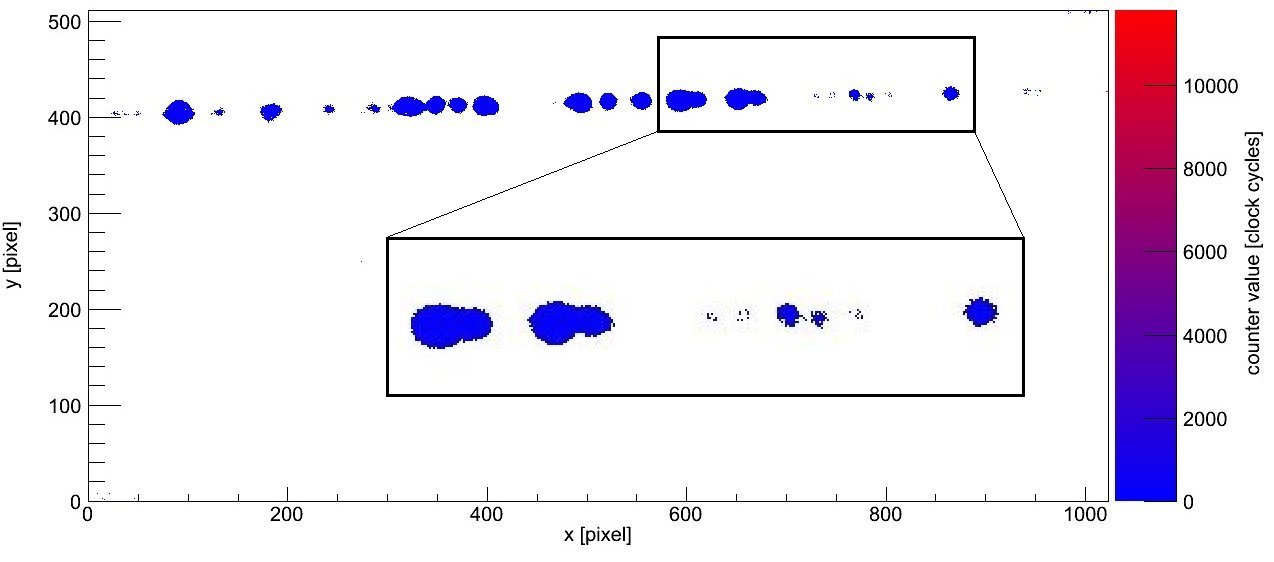}
\caption{Online event display of an event from the GEM module.}
\label{fig:eventdisplay2}
\end{figure}

A first preliminary analysis has been started using the MarlinTPC~framework~\cite{Marlin}. For the track reconstruction a two dimensional Hough transformation and a track fit was used. First results are presented here and discussed in full detail in \cite{menzen}. 
In figure~\ref{fig:eventdisplay1} and \ref{fig:eventdisplay2} online event display images of tracks for the two modules are shown. For the InGrid module, a double track event with clearly visible primary electron signals can be seen, whereas for the GEM module the track consists of several big blobs with less information about the number of primary electrons. In case of a GEM, several primary electrons end up in a single hole of the top GEM. Moreover the signal will spread out in the triple GEM stack and along the drift inside the stack towards the Timepix chips and create a large charge deposition. If the ionisation density of the charged particle is high, these depositions may overlap. Due to the characteristics of the InGrid detector (see section\ref{sec:InGrid}) the primary electrons are still visible. For the first analysis, a dataset of a z-scan with $\mathrm{E_{Drift}} = \SI[per-mode=symbol]{230}{\volt\per\centi\meter}$ was chosen. The transverse diffusion for this field configuration was calculated with MAGBOLTZ~\cite{magboltz} to $\mathrm{D_T(B = 0 T)} \approx \SI[per-mode=symbol]{310}{\micro\meter\per\sqrt{\centi\meter}}$ and $\mathrm{D_T(B = 1 T)} \approx \SI[per-mode=symbol]{100}{\micro\meter\per\sqrt{\centi\meter}}$.

\begin{figure}[tbp] % figures (and tables) should go top or bottom of
                    % the page where they are first cited or in
                    % subsequent pages
\centering
\includegraphics[width=1.0\textwidth]{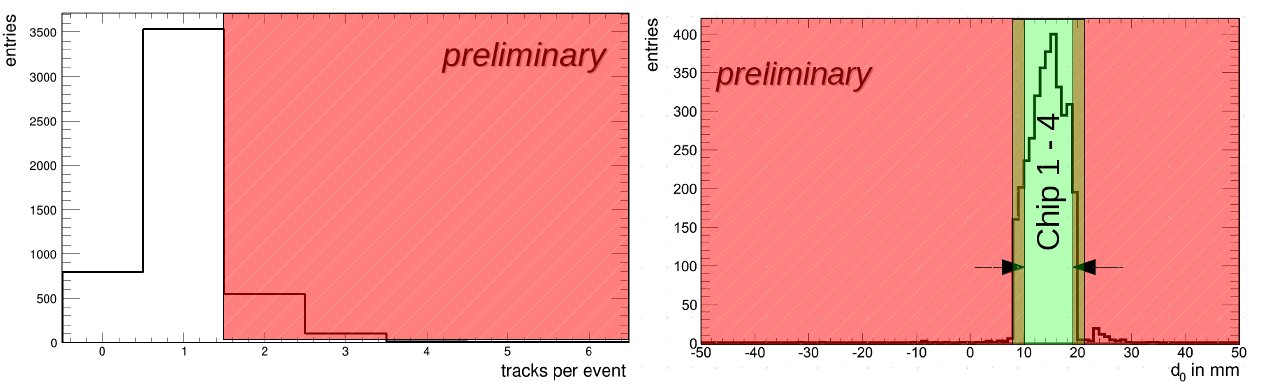}
\caption{Cuts applied for the preliminary analysis: events with more than one track were rejected (left), only tracks from the center of the lower chip row (green, chips 1-4) were accepted (right). The x-axis parameter $\mathrm{d_0}$ is the minimal distance in the x-y-plane of the track to the coordinate origin, see figure~\protect\ref{fig:trackpara}.}
\label{fig:cuts}
\end{figure}

\begin{figure}[tbp] % figures (and tables) should go top or bottom of
                    % the page where they are first cited or in
                    % subsequent pages
\centering
\includegraphics[width=1.0\textwidth]{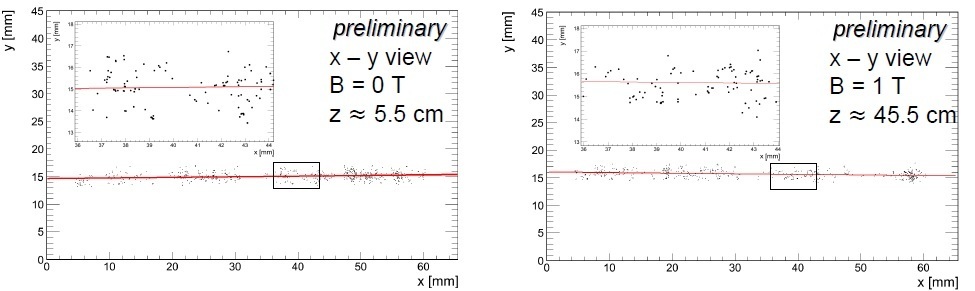}
\caption{Reconstructed tracks for $\mathrm{B} = \SI{0}{\tesla}$(left) and $\mathrm{B} = \SI{1}{\tesla}$ (right) with similar transverse diffusion. Note the different z-positions.}
\label{fig:recoTracks}
\end{figure}

\paragraph{Cuts}
Due to a rarely occurring bit shift error in the readout system $\approx 4\%$ of the hits are outside the shutter window. They are physically not meaningful. One fourth of this originated from non-hit pixels, another fourth from pixels hit during the shutter window. From the recorded data only the physically meaningful hits from within the shutter window were accepted.
For a simple analysis, cuts were applied to select single straight tracks, see figure~\ref{fig:cuts}. First of all, only tracks with more than 200~hits were accepted. This excludes track segments and $\mathrm{\delta}$-electrons. Next, events with more than two tracks were rejected to simplify the analysis. Finally, tracks were excluded that are too close to the lower and upper borders of the lower chip row (chips 1-4) such that some primary electrons could have diffused out of the sensitive area. Since the residuals of the hits on a track depend on the distance the primary electrons have to drift towards the anode, this cut depends on the z-position of the track. The cut was set to a distance of 3~$\mathrm{\sigma}$ of the expected transverse diffusion width. Chips 1-4 were choosen, as the beam was focussed on this lower chip row for data taking.\\
Figure~\ref{fig:recoTracks} shows two reconstructed tracks in the x-y~plane. One for $\mathrm{B} = \SI{0}{\tesla}$ and another for $\mathrm{B} = \SI{1}{\tesla}$. The track with $\mathrm{B} = \SI{1}{\tesla}$ originates from a eight times larger distance in z-direction. The suppression of the transverse diffusion is clearly visible, as both tracks have approximately the same width. Another remarkable fact can be seen in the enlarged part of the figure: The individual hits of the primary electrons are clearly visible. Even on the short track length of \SI{8}{\milli\meter} there are $O(100)$ track points.

\begin{figure}[tbp] % figures (and tables) should go top or bottom of
                    % the page where they are first cited or in
                    % subsequent pages
\centering
\includegraphics[width=.6\textwidth]{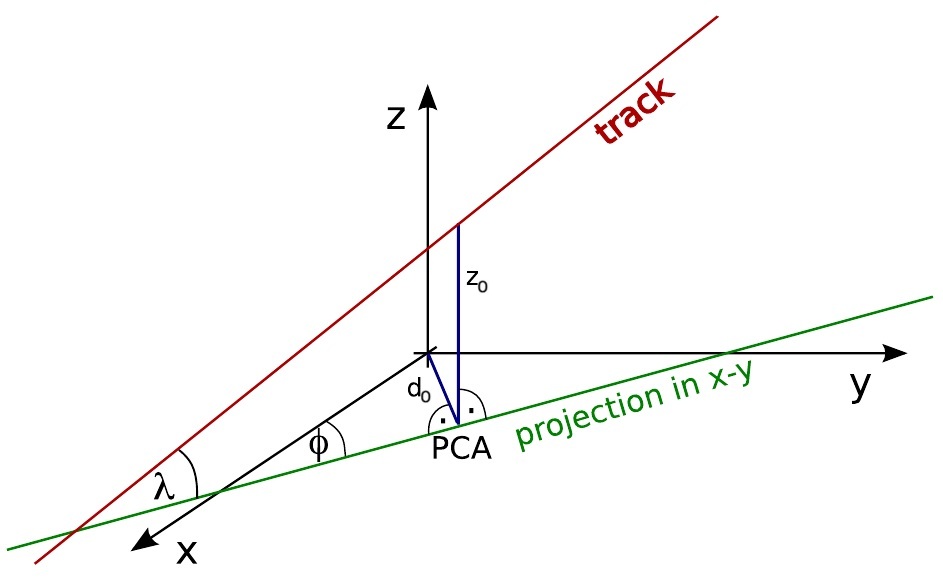}
\caption{Illustration of the track parameters $z_r0$, $d_0$, $\lambda$, and $\phi$, \cite{simone}}
\label{fig:trackpara}
\end{figure}

\begin{figure}[tbp] % figures (and tables) should go top or bottom of
                    % the page where they are first cited or in
                    % subsequent pages
\centering
\includegraphics[width=1.0\textwidth]{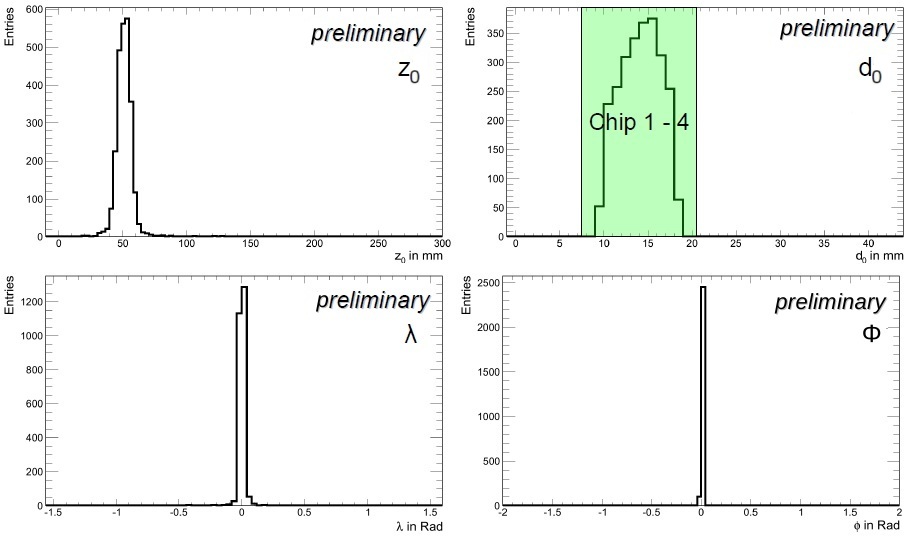}
\caption{Distributions of the track parameters $\mathrm{z_0}$, $\mathrm{d_0}$, $\mathrm{\lambda}$, and $\mathrm{}\phi$. for a run with $\mathrm{B} = \SI{0}{\tesla}$ and $z \approx \SI{5,58}{\centi\meter}$.}
\label{fig:controllplots}
\end{figure}

\paragraph{Track parameters}
The geometrical parameters used in the track reconstruction are shown in figure~\ref{fig:trackpara}. For the test beam data, the origin of the coordinate system was chosen to be outside the eight InGrids on the octoboard carrier. A reconstructed track can be described by four parameters: $\mathrm{d_0}$ is the minimum distance of the track projection to the origin in the readout plane (note the point of closest approach PCA) and $\mathrm{\phi}$ the angle towards the x-axis in this plane. $\mathrm{\lambda}$ is the angle of the track towards this plane and $\mathrm{z_0}$ the distance in z-direction from the PCA to the track.\\
In figure~\ref{fig:controllplots}, the measured track parameters are shown for a dataset with $\mathrm{B} = \SI{0}{\tesla}$. Since for the first analysis a simple dataset was chosen, the parameters can easily be interpreted: the tracks originate from a single z-position around \SI{50}{\milli\meter} ($\mathrm{z_0}$ plot) and are concentrated on chip 1-4 ($d_0$ plot). Note that the origin of the coordinate system is outside the sensitive area and chips 1-4 range from $y \approx \SI{7,5}{\milli\meter}$ to $y \approx \SI{21,5}{\milli\meter}$. This is reflected by $d_0$ as the tracks are almost parallel to the x-axis ($\mathrm{\phi}$ plot). The tracks are also parallel to the x-y-plane that is the readout plane.

\begin{figure}[tbp] % figures (and tables) should go top or bottom of
                    % the page where they are first cited or in
                    % subsequent pages
\centering
\includegraphics[width=1.0\textwidth]{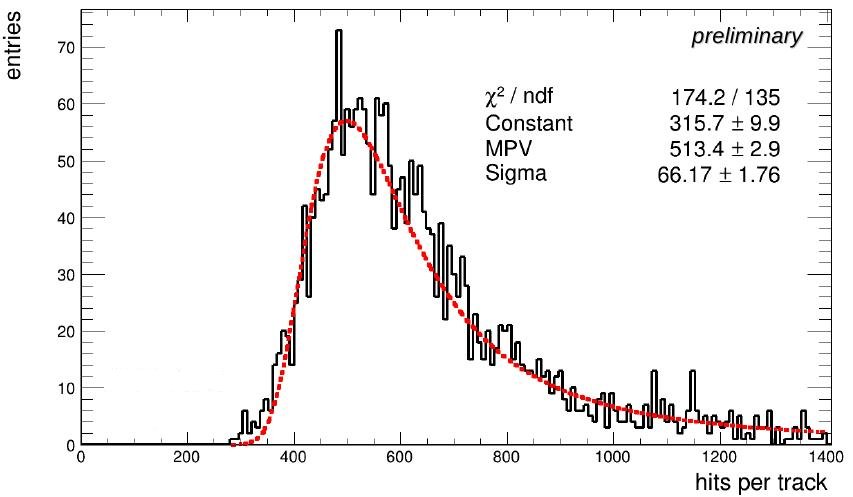}
\caption{Hits per track for a run with $\mathrm{B} = \SI{0}{\tesla}$ and $\mathrm{z} \approx \SI{5,58}{\centi\meter}$ with a Landau distribution fitted (dashed red line).}
\label{fig:hitspertrack}
\end{figure}

\paragraph{Hits per track}
The number of hits per track for a specific run with $\mathrm{B} = \SI{1}{\tesla}$ at a z-position of approximately \SI{5,58}{\centi\meter} is shown in figure~\ref{fig:hitspertrack}. The distribution can be described by a Landau function, as expected. The most probable value for this particular run gives a value of 513~hits. Tracks with less then 200~hits were not accepted by the analysis. Tracks with more than 1400~hits include regions with high densities of primary ionisations most likely caused by $\mathrm{\delta}$ electrons  not leaving the track perimeter. Events with "real" $\mathrm{\delta}$ electrons were rejected by this analysis, as those are usually detected as a second track.\\
As the beam crossed four chips, the length of the tracks is approximately \SI{5,6}{\centi\meter}. The number of hits per track length is around \SI[per-mode=symbol]{90}{hits\per\centi\meter} which is a reasonable value for T2K gas. Considering the inactive area at the edges of the chips, this number has to be corrected slightly upwards.

\begin{figure}[tbp] % figures (and tables) should go top or bottom of
                    % the page where they are first cited or in
                    % subsequent pages
\centering
\includegraphics[width=1.0\textwidth]{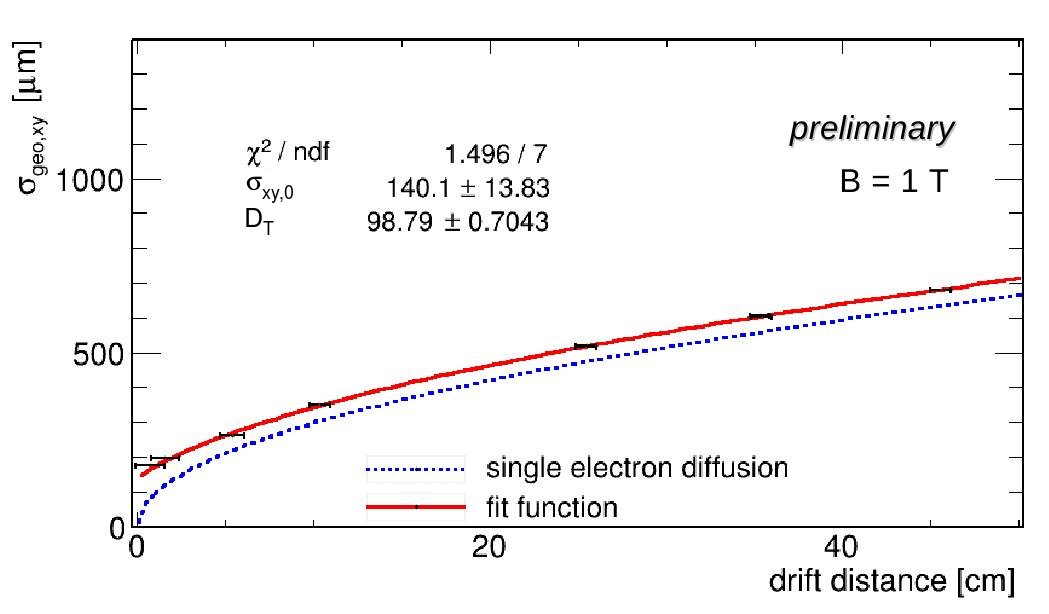}
\caption{Observed transversal spatial resolution for $\mathrm{B} = \SI{1}{\tesla}$ and $E_{Drift} =$~\SI[per-mode=symbol]{230}{\volt\per\centi\meter} fitted by the equation~~\protect\ref{eq:xyresol} and compared to the theoretical limit given by the single electron diffusion (dashed line).}
\label{fig:xyresol}
\end{figure}

\paragraph{Transverse spatial resolution}
Most of the pixels hit in an InGrid detector are activated by single primary electrons. Hence, ideally the transverse spatial resolution follows the single electron diffusion limit. In reality the performance is reduced by detector imperfections as for example misalignment and field distortions. In figure~\ref{fig:xyresol}, the transverse spatial resolution is shown together with the single electron diffusion limit. The resolution was measured at different z-positions (z-scan) and a function fitted to the measurement that is given by
\begin{equation}
\label{eq:xyresol}
\mathrm{\sigma_{geo,xy}(z)=\sqrt{\sigma_{xy,0}^2+D_T^2\cdot{z}}},
\end{equation}
where $\mathrm{\sigma_{geo,xy}}$ is calculated by the geometric mean method \cite{magboltz}:
\begin{equation}
\mathrm{\sigma_{geo,xy}=\sqrt{\sigma_{N-1,xy}\cdot{\sigma_{N,xy}}}}.
\end{equation}
$\mathrm{\sigma_{N,xy}}$ is the mean distance in the x-y plane of a hit to a track, fitted by using all of the hits belonging to the track, whereas $\mathrm{\sigma_{N-1,xy}}$ is the mean distance in the x-y plane of a hit to a track, fitted by using all hits belonging to the track but the one under investigation.
We can see from the fit parameters that the transverse diffusion constant was measured to be 
$\mathrm{D_T(B = 1 T)} = (99 \pm 1)\SI{}{\micro\meter\per\sqrt{\centi\meter}}$ which is in agreement with the expectation. The intrinsic detector resolution is given by $\mathrm{\sigma_{xy,0}}$, the extrapolation of the fit to $\mathrm{z}=\SI{0}{\centi\meter}$. This is expected to be the pixel pitch divided by $\sqrt{12}$ which is $\mathrm{\sigma_{xy,0,theo}} = \SI{15,9}{\micro\meter}$. However, the measured intrinsic detector resolution is $\mathrm{\sigma_{xy,0}} = (140 \pm 14) \SI{}{\micro\meter}$ causing the gap between the single electron diffusion and the fit. This value is largely dominated by field distortions originating from the chip boundaries as already shown in \cite{fielddistort}. In the ongoing analysis, these field distortions are clearly visible and will be further investigated.

\section{Conclusion and outlook}
The combination of micropattern gaseous detectors with pixelated readout chips was first investigated ten years ago and is now at a stage where those detectors are applicable in particle physics experiments. The most promising device is an integrated Micromegas on top of the Timepix ASIC, a so called InGrid. This fine grained high resolution detector can detect individual electrons and is now available in mass production. For larger scales, as for example a pixel TPC at the ILC, a new readout based on the Scalable Readout System is under development. A setup of eight InGrids read out by the SRS was tested at the large TPC prototype at DESY using T2K gas. Preliminary results show the potential of this readout technique. The effects limiting the intrinsic detector resolution will be studied in a more detailed analysis.\\
The Timepix3~\cite{Timepix3}, that is now available will provide new features as for example the  simultaneous measurement of charge and time or a higher sampling frequency.
On the readout side the SRS will be further improved. In an upcoming version one front end card will be able to host up to four octoboards. A system to support a 96 chip module is the short term goal, which would be a full LCTPC prototype module. This will demonstrate the feasibility of a large scale pixel TPC.

\acknowledgments

I would like to thank my collaborators from the LCTPC-pixel groups at CEA Sacly, DESY, NIKHEF, University of Bonn, LAL and University of Kiew. A special gratitude goes to Robert Menzen for the analysis and the support at the test beam.\\
The research leading to these results has received funding from the European Commission under the FP7 Research Infrastructure project AIDA, grant agreement no. 262025.

\end{document}